\documentclass[journal]{IEEEtran}
\usepackage{cite}
\usepackage{graphicx}
\usepackage{float}
\usepackage{comment}
\usepackage{caption}
\usepackage{subcaption}
\usepackage{amsmath}
\usepackage{hyperref}
\usepackage{soul}
\usepackage[dvipsnames]{xcolor}
\usepackage{tikz}
\ifCLASSINFOpdf
\else
\fi
\begin{document}
\title{Long-distance high-precision and high-sensitivity time delay sensing based on fiber optic weak measurements}
\author{Wei-Qian Zhao, 
Zi-Fu Su,
 Ya-Fei Yu,
and Jin-Dong Wang
\thanks{
W.-Q. Zhao,  and Z.-F. Su are with Guangdong Provincial Key Laboratory of Nanophotonic Functional Materials and Devices, School of Optoelectronic Science and Engineering,  South China Normal University, Guangzhou 510006, China
(email: wqzhaoqua@m.scnu.edu.cn; 2023022373@m.scnu.edu.cn).

Y.-F. Yu is with is with Guangdong Provincial Key Laboratory of Nanophotonic Functional Materials and Devices, School of Optoelectronic Science and Engineering,  South China Normal University, Guangzhou 510006, China, and also with Quantum Science Center of Guangdong-Hong Kong-Macao Greater Bay Area, Shenzhen-Hong Kong International Science and Technology Park,  NO.3 Binglang Road, Futian District, Shenzhen, Guangdong
(email: yuyafei@m.scnu.edu.cn).

 J.-D. Wang is with Guangdong Provincial Key Laboratory of Quantum Engineering and Quantum Materials, School of Optoelectronic Science and Engineering, South China Normal University, Guangzhou 510006, China, is also with Guangdong Provincial Key Laboratory of Nanophotonic Functional Materials and Device,  Guangdong Basic Research Center of Excellence for Structure and Fundamental Interactions of Matter, 
School of Optoelectronic Science and Engineering, 
South China Normal University, Guangzhou 510006, China
(email: wangjingdong@m.scnu.edu.cn).
}
%\thanks{E-mail: lkox@dtu.dk, smasw@dtu.dk}
\thanks{Manuscript received xx}}
\markboth{Journal of \LaTeX\ Class Files,~Vol.~XX, No.~X, JANUARY~2024}
{Shell \MakeLowercase{\textit{et al.}}: Bare Demo of IEEEtran.cls for IEEE Journals}

\maketitle

\begin{abstract}
In fiber optic sensing, time delays induced by polarization mode dispersion can distort signals in systems relying on phase or intensity variations for measurement, degrading performance, especially in long distance, high-precision applications. To address this challenge, we propose a weak measurement-based scheme using intensity contrast ratio for high-precision, high-sensitivity fiber optic delay estimation under large inherent time delays. We demonstrate that a narrower light source bandwidth enhances the effective sensing distance for high-sensitivity measurements. Our results show that, even with large inherent time delays, the measurement precision and sensitivity remain comparable to those of biased weak measurement, enabling detection of time delay variations at the attosecond level, corresponding to a 25.5 Pa water pressure change. The scheme is also robust against fiber misalignment errors, offering a novel solution for long-distance distributed fiber-optic sensing and broadening the applications of weak measurement techniques.
\end{abstract}

\IEEEpeerreviewmaketitle
\section{Introduction}

\IEEEPARstart{T}{he} Optical fiber sensing technology plays a crucial role in precision measurements, particularly in time delay estimation, vibration monitoring, and pressure sensing. However, traditional fiber optic sensing systems often face issues such as dispersion noise, signal attenuation due to device losses, and insufficient sensitivity  \cite{FST1, FST2, FST3, FST4, FST5}, which limit their performance in high-precision applications.

In recent years, the development of weak measurement (WM) theory and technology \cite{PhysRevLett.60.1351} has led to the emergence of WM-based fiber optic sensing as a novel approach to enhancing sensor performance. By exploiting the weak value amplification (WVA) effect \cite{PhysRevA.41.11}, WM technology not only significantly improves the sensitivity of optical fiber sensors but also enables high-precision sensing with the benefits of low cost, small size, and high stability \cite{PhysRevLett.93.203902, 8691849, Luo:20, huang2021design, liu2022ultra}.

Although WM techniques can significantly improve sensing performance \cite{PhysRevA.100.012109, Xu_2024, luo2019simultaneously, PhysRevA.97.033851, PhysRevApplied.18.044031}, their effectiveness depends on the appropriate combination of pre- and post-selection to achieve the WVA effect \cite{PhysRevLett.105.010405, PhysRevLett.111.033604}. However, this amplification also requires satisfying weak interaction conditions. Therefore, several challenges remain in practical fiber optic sensing applications. Firstly, birefringence cross-talk in optical fibers causes amplitude type noise , which significantly reduces the sensitivity of the sensor. While this noise can be partially suppressed by introducing biased weak measurement (BWM) theory and selecting appropriate detection methods, but the sensing performance still lags behind that of free-space optical systems \cite{wang2022robustness}. Secondly, when using the BWM theory, if the light source bandwidth is large, the pre-selected polarization state undergoes decoherence due to the large inherent time delays induced by polarization mode dispersion (PMD) in the optical fiber. This makes it difficult to apply WM in frequency-domain analysis for optical fiber time delay estimation.

To address this issue, existing studies in fiber optic WM applicaations, such as hydrostatic pressure sensing \cite{8691849} and vibration sensing \cite{Luo:20}, have employed optical fibers with identical lengths and vertically aligned slow axes for time delay compensation. Additionally, spectral splitting techniques have been used for detection. Although this approach has yielded promising results, it increases system complexity and cost, and requires precise control over fiber length. Moreover, it necessitates a broad bandwidth light source to meet the resolution conditions for spectral splitting \cite{wang2022robustness}. However, the broader spectrum exacerbates the impact of fiber PMD on the signal, making the implementation of long-range, distributed fiber-optic sensing more difficult.

In this work, we propose a WM-based scheme for high-precision fiber-optic delay estimation under large inherent time delays.  This method combines the theory of intensity contrast ratio \cite{qiu2017precision} in WM with fiber optic sensing, enabling precise estimation of time delay variations, even in the presence of large inherent time delays.  This method not only effectively mitigates the impact of PMD on the signal, thereby enhancing the sensing range, but its main advantage lies in the ability to adjust the post-selection angle based on the intensity of the fiber-optic output, enabling precise control over the estimation process. The proposed scheme offers promising potential for a wide range of applications in the field of fiber optic sensing.

\section{Theoretical Framework}
\label{sec:II}
\subsection{WM-based scheme under large inherent time delays}

The standard WM protocol generally involves four steps \cite{RevModPhys.86.307}: Pre-selection, where the initial state $|\varphi_{i} \rangle$ of the system is prepared; Weak interaction, during which the time delay to be measured is introduced, represented by the evolution operator $\hat{U}$; Post-selection, where the evolved state of the system is projected onto a chosen post-selected state $|\varphi_{f} \rangle$; Detection, where the required measurement information is obtained from the final state $|\phi_{f} \rangle$ of the pointer.

In the proposed scheme, we aim to measure the time delay variation $\Delta \tau$ over long fiber distance where large inherent time delays $\tau_{0}$ are present, and the measurement principle is shown in Fig. 1.

\begin{figure}[t]
    \centering
\includegraphics[width=\linewidth,keepaspectratio]{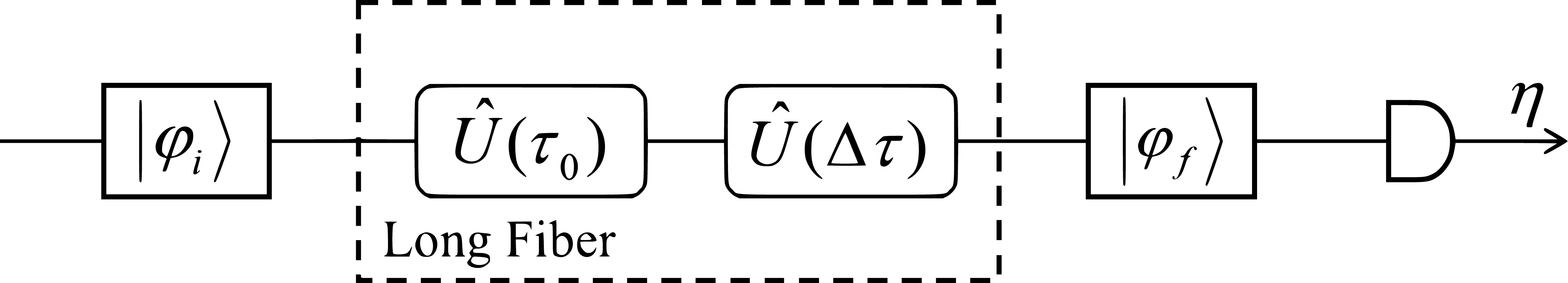}
    \caption{Schematic diagram of the WM-based scheme for high-precision fiber optic delay estimation under large inherent time delays. Here, $\hat{U} \left( \tau_{0} \right)$ represents the significant inherent time delay induced by the evolution in the long-distance optical fiber, and $\hat{U} \left( \Delta \tau \right)$ denotes the delay variation caused by environmental perturbations affecting the fiber.}
    \label{fig:1}
\end{figure}

The pre-selected state of the system is $|\varphi_{i} \rangle =\frac{1}{\sqrt{2}} \left( |H\rangle +|V\rangle \right)$, and the post-selected state is $|\varphi_{f} \rangle =\frac{1}{\sqrt{2}} \left( e^{-i\epsilon}|H\rangle -e^{i\epsilon}|V\rangle \right)$, where $|H\rangle$ and $|V\rangle$ represent the horizontal and vertical polarization states, respectively, and correspond to the two orthogonal mode components of the optical fiber.

The pointer is characterized by the spectral distribution of a monochromatic light source, with an initial state given by $|\phi \rangle =\int d\omega f\left( \omega \right) |\omega \rangle$, where $f\left( \omega \right) =\left( \pi \sigma^{2} \right)^{-\frac{1}{4}} \text{exp} \left[ -\frac{\left( \omega -\omega_{0} \right)^{2}}{2\sigma^{2}} \right]$ is the spectral distribution function, $\sigma^{2} =\langle \omega^{2} \rangle -\langle \omega \rangle^{2}$ represents the light source bandwidth, and $\omega_0$ denotes the central frequency of the initial spectrum.

The system and pointer interact with each other through the long-distance optical fiber, resulting in a joint state expressed as:
\begin{equation}
\begin{aligned}
|\Psi \rangle &= \hat{U} \left( |\varphi_{i} \rangle \otimes |\phi \rangle \right) \\
&= \frac{1}{\sqrt{2}} \int \left( e^{-i\omega \tau} |H\rangle + e^{i\omega \tau} |V\rangle \right) d\omega |\omega \rangle.
\end{aligned}
\end{equation}

The interaction $\hat{U}$, which between the system and pointer is described by $\hat{U} =e^{-i\tau \omega \hat{A}}$, where $\hat{A} =|H\rangle \langle H|-|V\rangle \langle V|$ is the observable operator for the system \cite{PhysRevLett.115.120401, PhysRevLett.110.083605}. Inside the optical fiber, the interaction induces two types of time delay, as illustrated in Fig 1. The first component arises from the birefringence of the optical fiber, resulting in a large inherent time delays $\tau_{0}$, which is proportional to the length of the fiber. The second component corresponds the small time delay variation $\Delta \tau$ caused by external disturbances affecting the sensing fiber during the measurement process \cite{8691849, Luo:20}. The total time delay within the optical fiber during measurement can be expressed as $2\tau =2\tau_{0} +2\Delta \tau$, and $\tau_{0} >>\Delta \tau$.

The joint state $|\Psi \rangle$ is projected onto the post-selected state $|\varphi_{f} \rangle$, yielding the final normalized state of the pointer:

\begin{equation}
\begin{aligned}
|\phi_{f} \rangle &= \langle \varphi_{f} |\hat{U} |\varphi_{i} \rangle |\phi \rangle /\sqrt{P_{f}} \\
&= \frac{1}{\sqrt{P_{f}}} \int d\omega f\left( \omega \right) \sin \left( \omega \tau -\varepsilon \right) |\omega \rangle,
\end{aligned}
\end{equation}
where $P_{f}$ represents the probability of successful post-selection, ensuring the pointer final state is properly normalized,

\begin{equation}
\begin{aligned}
P_{f} &= \left| \langle \varphi_{f} |\hat{U} |\varphi_{i} \rangle |\phi \rangle \right|^{2} \\
&= \frac{1}{2} \left[ 1 - e^{-\sigma^{2} \tau^{2}} \cos \left( 2\omega_{0} \tau - 2\epsilon \right) \right].
\end{aligned}
\end{equation}

In our scheme, we use the output optical power after post-selection as a pointer, assuming that the initial input optical power is $I_{in}$ and the output optical power after post-selection is $I_{out}=I_{in}P_f$. At the initial stage of the experiment, no additional perturbation is applied to the sensing fiber so that the time delay in the sensing fiber is generated by birefringence, i.e., $\tau =\tau_{0}$, and then the post-selection angle is adjusted so that the output light intensity reaches the minimum value of $I_{min}$. At this time, the post-selection angle is $\epsilon =\epsilon_{0}$, 
\begin{equation}
I_{min}=\frac{1}{2} I_{in}\left( 1-e^{-\sigma^{2} \tau_{0}^{2}} \right) ,
\end{equation}
and satisfies $\omega_{0} \tau_{0} -\epsilon_{0} =k\pi ,\  k=0,\  1,\  2......$
Subsequent fine-tuning of the posterior selection angle based on $I_{min}$ and rotation of $\Delta\epsilon$, resulting in a post-selection angle $\epsilon =\epsilon_{0} +\Delta \epsilon$. The output intensity obtained is denoted as $I_{0}$, which satisfies:
\begin{equation}
I_{0}=\frac{1}{2} I_{in}\left[ 1-e^{-\sigma^{2} \tau_{0}^{2}}\cos \left( 2\Delta \epsilon \right) \right].
\end{equation} 

After that, a disturbance is applied to the optical fiber, causing a time delay variations, i.e., $\tau=\tau_0+\Delta\tau$. The resulting output intensity is denoted as $I_1$, which satisfies:
\begin{equation}
I_{1}=\frac{1}{2} I_{in}\left[ 1-e^{-\sigma^{2} \tau_{0}^{2}}\cos \left( 2\Delta \epsilon -2\omega_{0} \Delta \tau \right) \right],
\end{equation} 
Finally, we obtain the intensity contrast ratio $\eta$, 
\begin{equation}
\eta =\frac{I_{0}-I_{1}}{I_{0}-I_{min}} \approx \frac{2\omega_{0} \Delta \tau}{\Delta \epsilon},
\end{equation} 
with the approximation condition $\omega_{0} \Delta \tau <<\Delta \varepsilon <<1,\  \sigma^{2} \tau^{2} <<1$. Thus, the fiber sensing time delay variation to be measured can be expressed as $\Delta \tau \approx \eta \Delta \varepsilon /2\omega_{0}$. The resolution of $\Delta \tau$ to be measured is determined by the magnitude of the post-selection angle and the minimum optical intensity variation detectable by the employed optical power meter.

In our scheme, the large inherent time delay $\tau_0$ in the sensing fiber does not imply a weak interaction. Instead, the $'weak'$ nature arises from the small time delay variation $\Delta \tau$ caused by external perturbations. We argue that $'weak'$ is not an inherent property of the system-pointer interaction but results from small changes in interaction strength due to external variations. This allows the existence of large inherent time delays, thus expanding the applicability and measurement range of weak measurement techniques.

\subsection{The effect of transmission distance on sensing sensitivity}

Here, we have discussed the bandwidth of the light source with respect to the effective transmission distance for high-sensitivity signals, defining $\kappa$ as the sensitivity of the intensity contrast ratio signal:
\begin{equation}
\begin{aligned}
\kappa &= \frac{\partial \eta}{\partial \Delta \tau} \\
&= \frac{-2e^{-\sigma^{2} \tau^{2}}\left[ \sigma^{2} \tau \cos \left( 2\gamma \right) +\omega_{0} \sin \left( 2\gamma \right) \right]}{1-e^{-\sigma^{2} \tau_{0}^{2}}\cos \left( 2\Delta \epsilon \right)},
\end{aligned}
\end{equation}
where $\tau=\tau_0+\Delta \tau$, $\gamma=\Delta \epsilon_0-\omega_0 \Delta \tau$, From this we find that the sensitivity $\kappa \propto \text{exp} \left( -\sigma^{2} \tau^{2} \right)$. Since $\sigma^{2} \tau^{2} <<1$, it becomes essential to balance the trade-off between the light source bandwidth and the scale of the inherent large delay in this scheme as illustrated in Fig. 2.

\begin{figure}[h]
    \centering
    \begin{tikzpicture}
        \node[anchor=south west,inner sep=0] (image1) at (0,0) {\includegraphics[width=8cm,height=6cm]{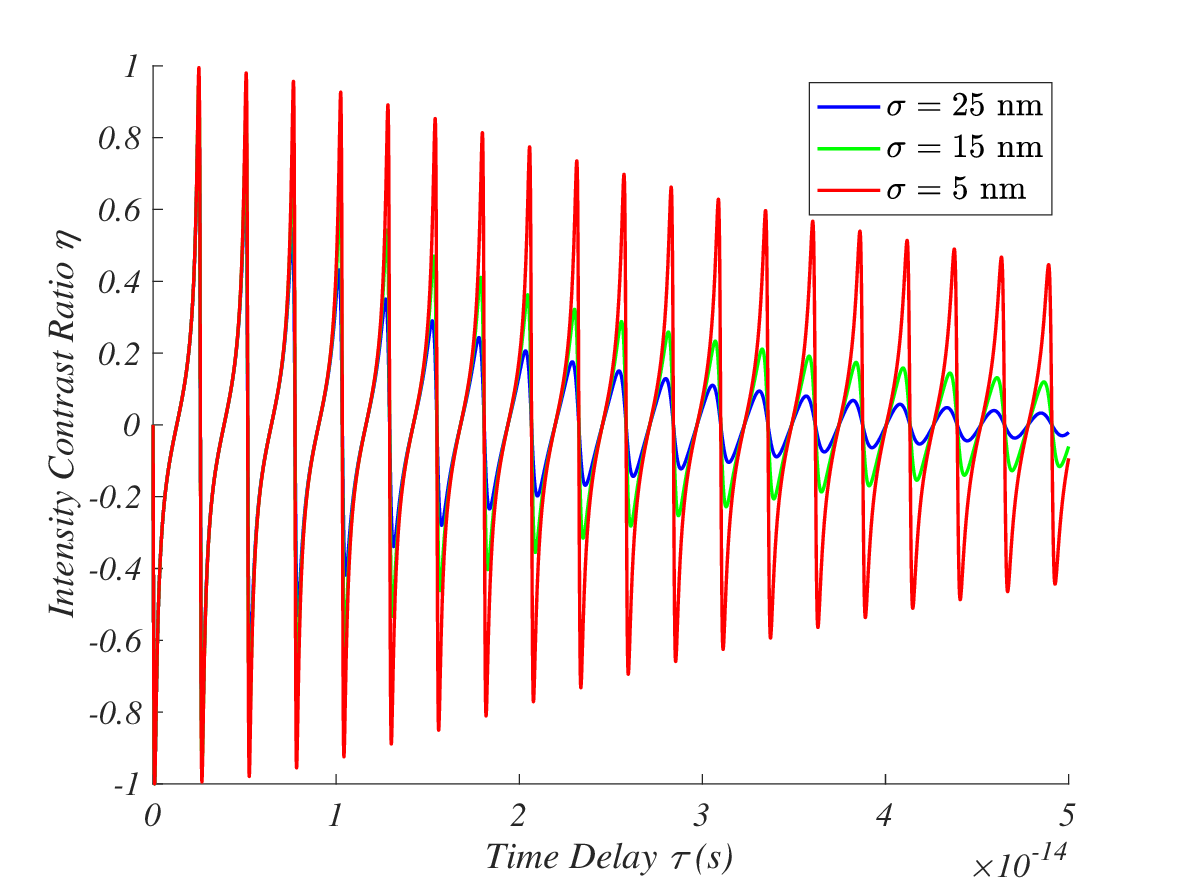}};
        \node at (0.4,5.5) {\textbf{(a)}};
    \end{tikzpicture}
    \begin{tikzpicture}
        \node[anchor=south west,inner sep=0] (image2) at (0,0) {\includegraphics[width=8cm,height=6cm]{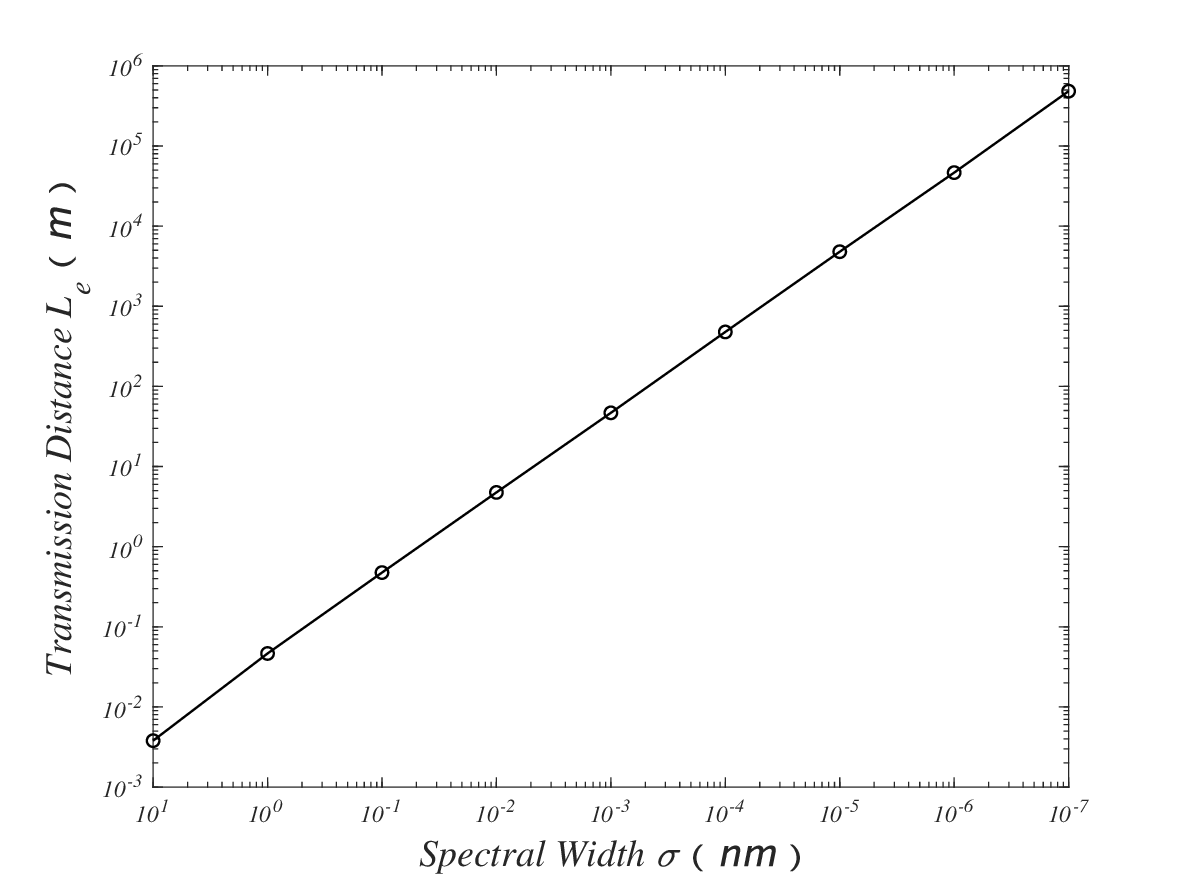}};
        \node at (0.4,5.5) {\textbf{(b)}};
    \end{tikzpicture}
    \caption{Discussion on the relationship between light source bandwidth $\sigma$ and the effective transmission distance for high-sensitivity signals $L_{e}$. (a) Intensity contrast ratio signal $\eta$ as a function of time delay $\tau$ for spectral widths of 5 nm, 15 nm, and 25 nm, represented by red, green, and blue solid lines, respectively. (b) The variation of the effective transmission distance for high-sensitivity signals $L_{e}$ with light source bandwidth $\sigma$ when the peak intensity contrast $\eta_{p}$ of its periodic variation drops to 0.9 due to the inherent time delay induced by increasing fiber length.}
    \label{fig:2}
\end{figure}

The peak intensity contrast ratio $\eta_{p}$ decreases more rapidly with increasing time delay $\tau$ as the spectral bandwidth $\sigma$ of the light source widens, reducing measurement sensitivity, as shown in Fig. 2(a). However, a narrower bandwidth allows the intensity contrast ratio $\eta$ to maintain higher peak values over a larger range of $\tau$, ensuring higher sensitivity. This behavior is governed by the amplitude term $\text{exp} \left( -\sigma^{2} \tau^{2} \right)$, which decays as $\sigma \tau$ increases. Since $\tau =2\pi BL/c$ grows with fiber length $L$, reducing $\sigma$ effectively extends the range of high-sensitivity measurements.

Theoretically, the smaller the value of $\sigma$, the more favorable it is for maintaining long-distance fiber sensor signals. Here, we define $L_e$ as the effective transmission distance for high-sensitivity signals, i.e., the fiber length corresponding to the time delay $\tau$ when the intensity contrast ratio peak value decreases to $\eta_e$ as $\tau$ increases, 
\begin{equation}
L_{e}=\frac{c\cdot \sqrt{\text{ln} \left( 1-\eta_{e} \right)^{-1}}}{2\pi B\sigma}.
\end{equation}

In order to ensure that the error in the variation of $\eta$ in $\Delta \tau \approx \eta \Delta \varepsilon /2\omega_{0}$ is less than one tenth of a percent, i.e., the accuracy is one order of magnitude smaller than the resolution, we discuss the variation curve of the effective transmission distance $L_e$ of the high-sensitivity signal when the intensity contrast peak $\eta_{e} =0.9$ is considered. For this, the spectral width range $\left[ 10^{-7},10 \right]$nm corresponds to the effective transmission distance $L_e$ of the high-sensitivity signal as shown in Fig. 2(b). It can be observed that the effective transmission distance $L_e$ for high-sensitivity signals increases exponentially as the light source bandwidth decreases. In other words, the narrower the bandwidth of the light source, the more favorable it is for long-distance fiber sensing signal transmission.

\subsection{The precision of parameter estimation with long transmission distance}

The quantum Fisher information (QFI) can be regarded as a measure of the responsiveness of a quantum state to parameter changes, reflecting the sensitivity of the quantum state to a given parameter \cite{e23030354}. In the WVA scheme, the weak interaction joint state $|\Psi \rangle$ of the system and the pointer contains all the metrological information of the parameter to be estimated, i.e., $Q_j$, while the post-selection step divides $Q_j$ into three parts \cite{PhysRevLett.125.080501}:
\begin{equation}
Q_{j}=P_{f}Q_{f}+P_{r}Q_{r}+F_{p},
\end{equation} 
where $P_{f}$ and $P_{r}$ represent the probabilities of successful and unsuccessful post-selection, respectively, while $Q_{f}$ and $Q_{r}$ correspond to the QFI of the pointer state upon successful and unsuccessful post-selection. Additionally, $F_{p}$ denotes the QFI associated with the probability distribution $\left\{ P_{f},P_{r} \right\}$.

For the joint state $|\Psi \rangle$ of the system and pointer after the weak interaction, the QFI $Q_{j}$ is expressed as:
\begin{equation}
Q_{j} = 4\left[ \frac{d\langle \Psi |}{d\tau} \frac{d|\Psi \rangle}{d\tau} -\left| \frac{d\langle \Psi |}{d\tau} |\Psi \rangle \right|^{2} \right] 
= 4\sigma^{2} + 4\omega_{0}^{2}.
\end{equation}

After the post-selection operation, the QFI $Q_{f}$ of the final pointer state $|\phi_{f} \rangle$ can be expressed as:
\begin{align}
Q_{f} &= 4\left[ \frac{d\langle \phi_{f} |}{d\tau} \frac{d|\phi_{f} \rangle}{d\tau} -\left| \frac{d\langle \phi_{f} |}{d\tau} |\phi_{f} \rangle \right|^{2} \right] \notag \\
&= 4P_{f}^{-1} \int \omega^{2} \cos^{2} \left( \omega \tau - \varepsilon \right) f^{2}(\omega) d\omega \notag \\
&\quad - P_{f} \left[ \int \omega \sin \left( 2\omega \tau - 2\varepsilon \right) f^{2}(\omega) d\omega \right]^{2}.
\end{align}

Considering the post-selection success probability $P_{f}$, the QFI $Q_{WVA}$ about the parameter $\tau$ during the WVA process is given by:
\begin{equation}
\begin{aligned}
    Q_{WVA} &= P_{f} Q_{f} \\
    &= 4 \int \omega^{2} \cos^{2} \left( \omega \tau - \varepsilon \right) f^{2}(\omega) d\omega \\
    &\quad - P_{f}^{2} \left[ \int \omega \sin \left( 2\omega \tau - 2\varepsilon \right) f^{2}(\omega) d\omega \right]^{2}.
\end{aligned}
\end{equation}

From the above equation, it can be observed that when $\omega_{0} \tau -\varepsilon =0$, the QFI $Q_{WVA}$ after the post-selection operation reaches its maximum value.

\begin{figure}[ht]
    \centering
    \begin{tikzpicture}
        \node[anchor=south west,inner sep=0] (image1) at (0,0) {\includegraphics[width=9cm,height=4cm]{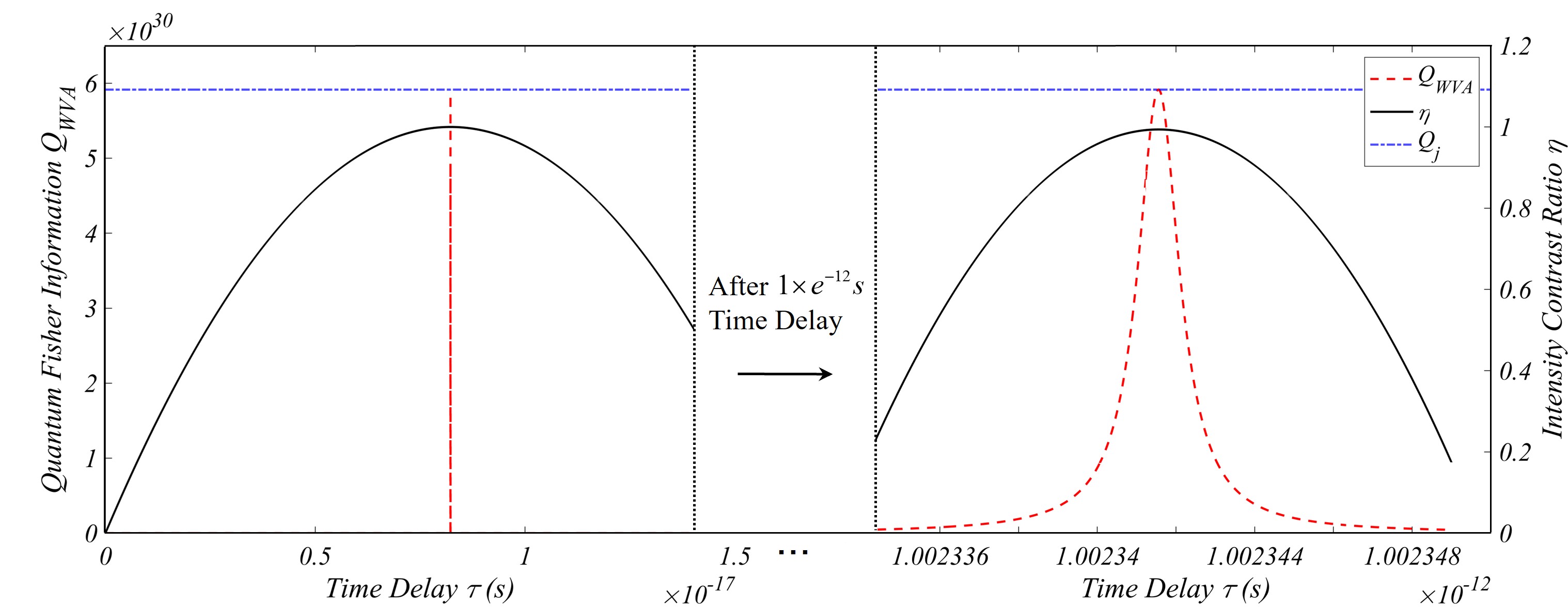}};
        \node at (0.2,3.8) {\textbf{(a)}};
    \end{tikzpicture}
    \begin{tikzpicture}
        \node[anchor=south west,inner sep=0] (image2) at (0,0) {\includegraphics[width=9cm,height=4cm]{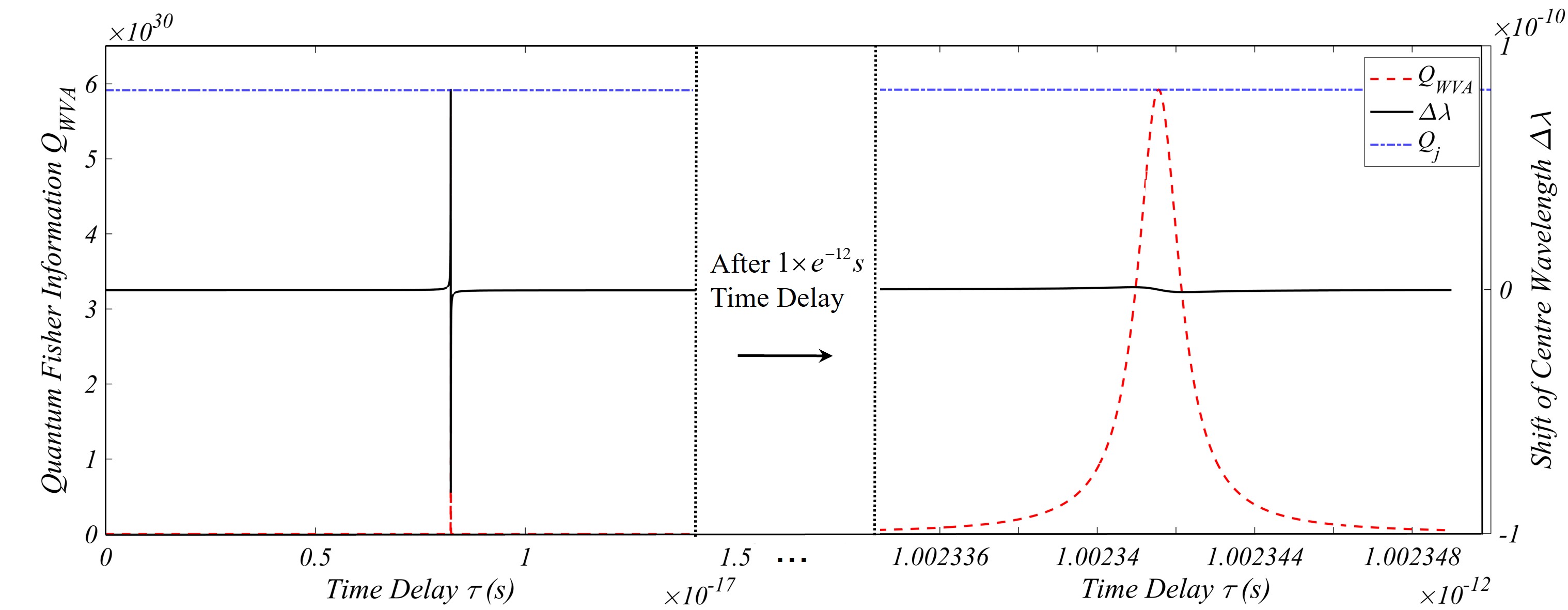}};
        \node at (0.2,3.8) {\textbf{(b)}};
    \end{tikzpicture}
    \caption{The distributions of QFI $Q_{WVA}$, intensity contrast ratio $\eta$, and shift of wavelength $\Delta\lambda$ under small and large time delay conditions. (a), the black solid line represents the intensity contrast $\eta$, the red dashed line represents the QFI distribution $Q_{WVA}$ after post-selection, and the blue dashed line represents the QFI $Q_{j}$ before post-selection. (b), the black solid line represents the wavelength shift $\Delta\lambda$, with the red and blue dashed lines matching those in (a). The time delay range between small and large delays differs by $10^{-12}$ orders of magnitude.}
    \label{fig:3}
\end{figure}

In Fig. 3, to provide a clearer representation of measurement sensitivity, we simulated the distributions of QFI $Q_{WVA}$, $Q_{j}$, intensity contrast ratio $\eta$, and shift of wavelength $\Delta\lambda$ as functions of the time delay $\tau$.   $Q_{j}$ represents the maximum QFI carried by the system, $Q_{WVA}$ represents the quantum Fisher information of the system after post-selection.

Fig. 3(a) shows that in the small time delay regime ($\sim 10^{-18}s$), the QFI distribution is narrow due to the limited light source bandwidth. Despite this, post-selection ensures the maximum QFI at specific points. In the larger time delay regime ($\sim 10^{-12}s$), PMD broadens the light source bandwidth and the QFI distribution, yet the peak QFI remains unchanged, maintaining high measurement precision. Additionally, the identical intensity contrast ratio $\eta$ distributions in both regimes confirm consistent measurement sensitivity, with the horizontal $\tau$ axis set to the range $15 as$ for direct comparison.

Fig. 3(b) highlights the wavelength shift behavior under the same coordinate setup. For a monochromatic light source, polarization mode dispersion reduces the wavelength shift as the sensing fiber length increases, limiting its utility in the linear region of WVA. However, the QFI in the linear region surpasses that in the nonlinear region, reflecting higher measurement precision and the advantage of biased weak measurements over standard ones \cite{PhysRevA.94.053843}. This WM-based scheme achieves high-precision, high-sensitivity time delay estimation by adjusting the post-selection angle via output light intensity feedback, maximizing QFI and enhancing sensitivity to intensity contrast changes.

\section{Experimental Demonstration}
\label{sec:III}
\subsection{Principle verification experiment}
To verify the feasibility of our WM-based scheme for high-precision fiber-optic delay estimation under large inherent time delays, we designed a principle verification experiment as shown in Fig. 4.

\begin{figure}[t]
    \centering
\includegraphics[width=\linewidth,keepaspectratio]{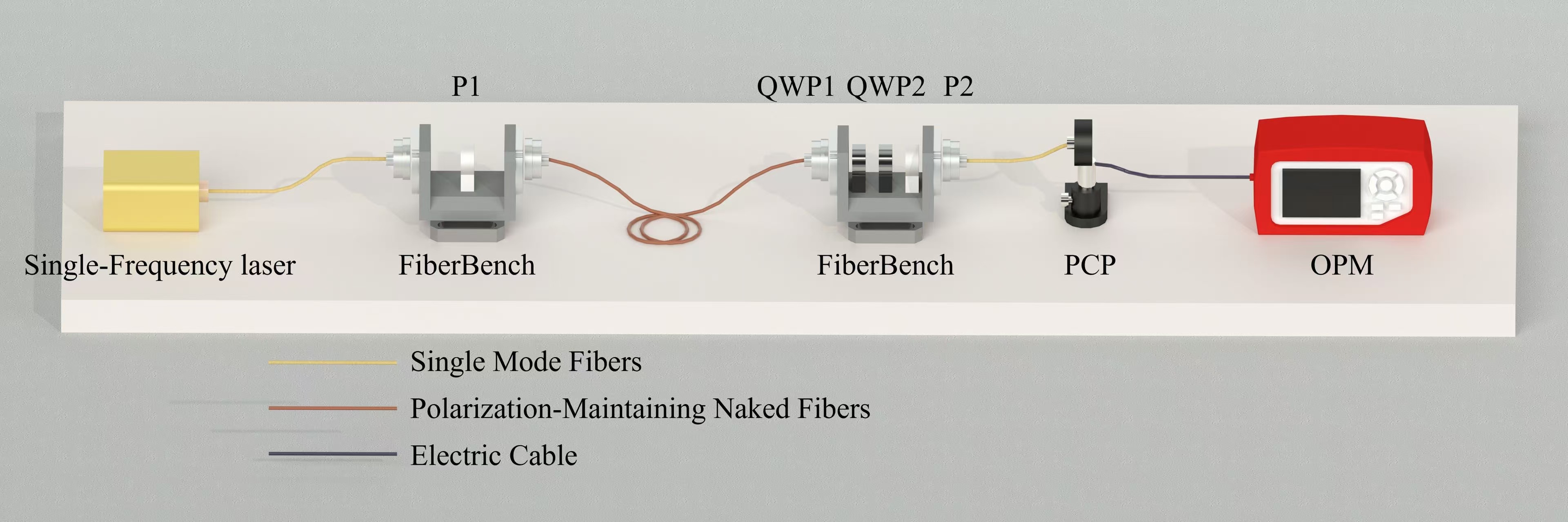}
    \caption{The experimental optical setup for the WM-based scheme for high-precision fiber optic delay estimation under large inherent time delays. The light source is a single-frequency laser. P1 and P2 are linear polarizers, employed for polarization control during the pre-selection and post-selection steps, respectively. QWP1 and QWP2 are quarter-wave plates, used to adjust the time delay and thereby control the output optical power. PCP denotes the photoelectric conversion probes, and OPM represents the optical power meter used for optical power detection.}
    \label{fig:4}
\end{figure}

$\textbf{Pre-selection}$: The light source (DenseLight, Tunable Single-Frequency Laser Box) is a monochromatic laser with a central wavelength of 1550$nm$ and a linewidth of 3.7$kHz$. A polarization-maintaining tail fiber is used to guide the linearly polarized light output from the source into the pre-selection free-space polarization control module (Thorlabs, FiberBench), achieving efficient fiber-space-fiber coupling. In the pre-selection module, a linear polarizer P1 is used for polarization generation, producing our pre-selection state: $|\varphi_{i} \rangle =\frac{1}{\sqrt{2}} \left( |H\rangle +|V\rangle \right)$. The basis vectors $H$ and $V$ are defined with respect to the fast and slow axes of the sensing fiber, ensuring that the prepared linearly polarized light at a $45^{\circ}$ angle to the slow axis of the sensing fiber has components along both the fast and slow axes.

$\textbf{Weak interaction}$: The linearly polarized light, prepared in the pre-selection step, is coupled into a 1-meter-long section of polarization-maintaining fiber (PMF) for the weak interaction step. The orthogonal components of the light align with the fast and slow axes of the PMF, introducing a large intrinsic time delay after passing through the fiber. This delay, $\tau_{0} =2\pi BL_{0}/c$, is related to the birefringence $B$ and length $L_{0}$ of the PMF, and in this experiment, it is on the order of \(10^{-12}\) s. With the chosen light source linewidth, a near-1 maximum intensity contrast ratio is achievable. The evolution operator $\hat{U} =e^{-i\tau \omega \hat{A}}$ describes the corresponding dynamics, where $\tau =\tau_{0} +\Delta \tau$.

$\textbf{Post-selection}$: This step is implemented using a polarization control module (FiberBench) consisting of two quarter-wave plates (QWP1 and QWP2) and a linear polarizer (P2) for feedback adjustment and post-selection. In this proof-of-concept experiment, QWP1 creates a small post-selection angle $\Delta \epsilon$, while QWP2 simulates small time delay variations $\Delta \tau$ in the sensing PMF. However, it is crucial to distinguish between  $\Delta \tau$ and the time delay induced by QWP2, as $\Delta \tau$ affects the amplitude term $e^{-\sigma^{2} \tau^{2}}$, whereas the time delay produced by QWP2 does not. Therefore, the system must be constrained to $\Delta \tau <<\tau_{0}$ to treat the delay produced by QWP2 as a small variation $\Delta \tau$.

$\textbf{Detection}$: Finally, the post-selected light beam is guided through a fiber pigtail into a oprical power meter (Thorlabs, PM100D) for power detection.

During the experiment, we first adjusted the optical axis directions of QWP1 and QWP2 in the post-selection module to minimize the output light intensity $I_{min}$, which served as the initial calibration condition. The measured residual light intensity $I_{min}$ during the experiment was on the order of $nW$, which is negligible compared to the $I_{0}$ and $I_{1}$ magnitudes of $\mu W$. Subsequently, QWP1 optical axis direction was finely adjusted to set the post-selection angle, and the sensing time delay variation $\Delta \tau$ was simulated by rotating QWP2. Each $0.1^{\circ}$ rotation of QWP2 corresponded to a time delay variation $\Delta \tau$ of $1.435\times 10^{-18}s$. The experimental results are shown in Fig. 5.

\begin{figure}[h]
    \centering
    \begin{tikzpicture}
        \node[anchor=south west,inner sep=0] (image1) at (0,0) {\includegraphics[width=8cm,height=6cm]{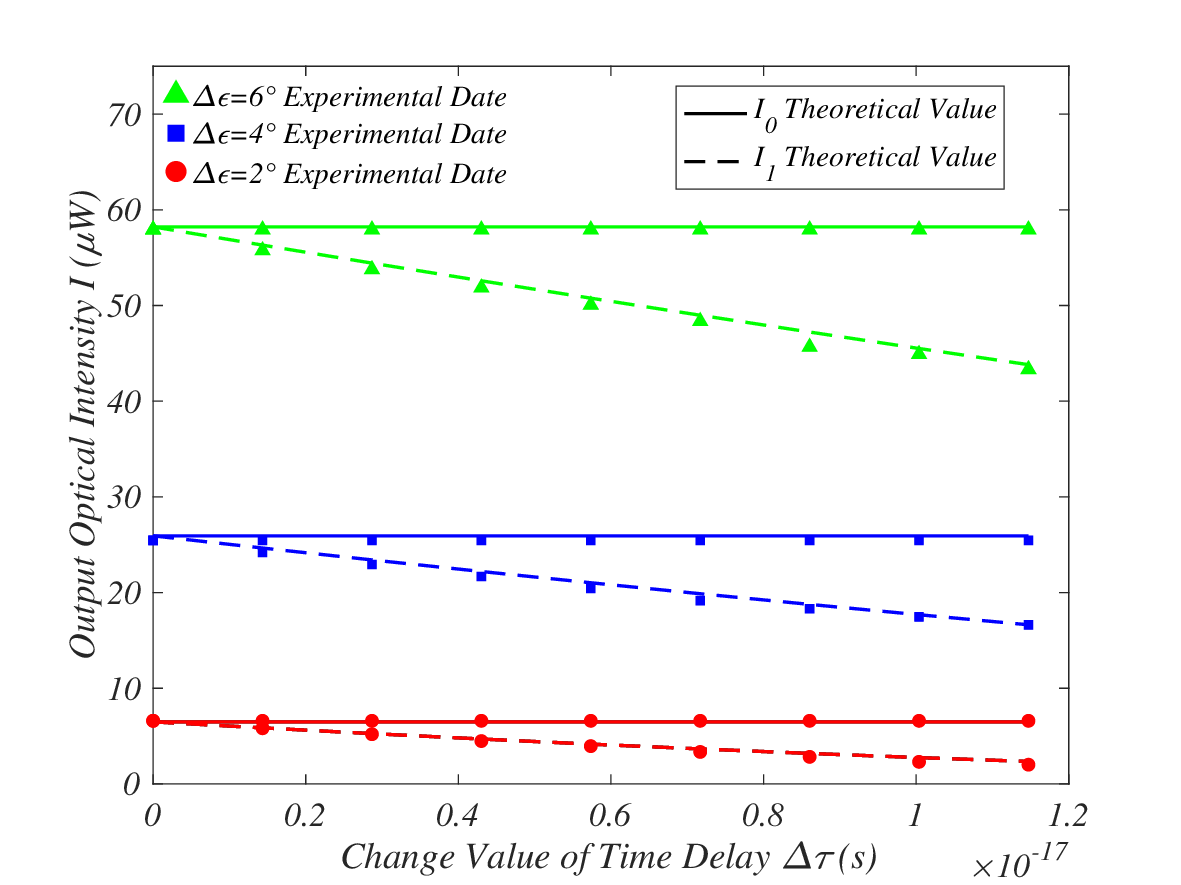}};
        \node at (0.4,5.5) {\textbf{(a)}};
    \end{tikzpicture}
    \begin{tikzpicture}
        \node[anchor=south west,inner sep=0] (image2) at (0,0) {\includegraphics[width=8cm,height=6cm]{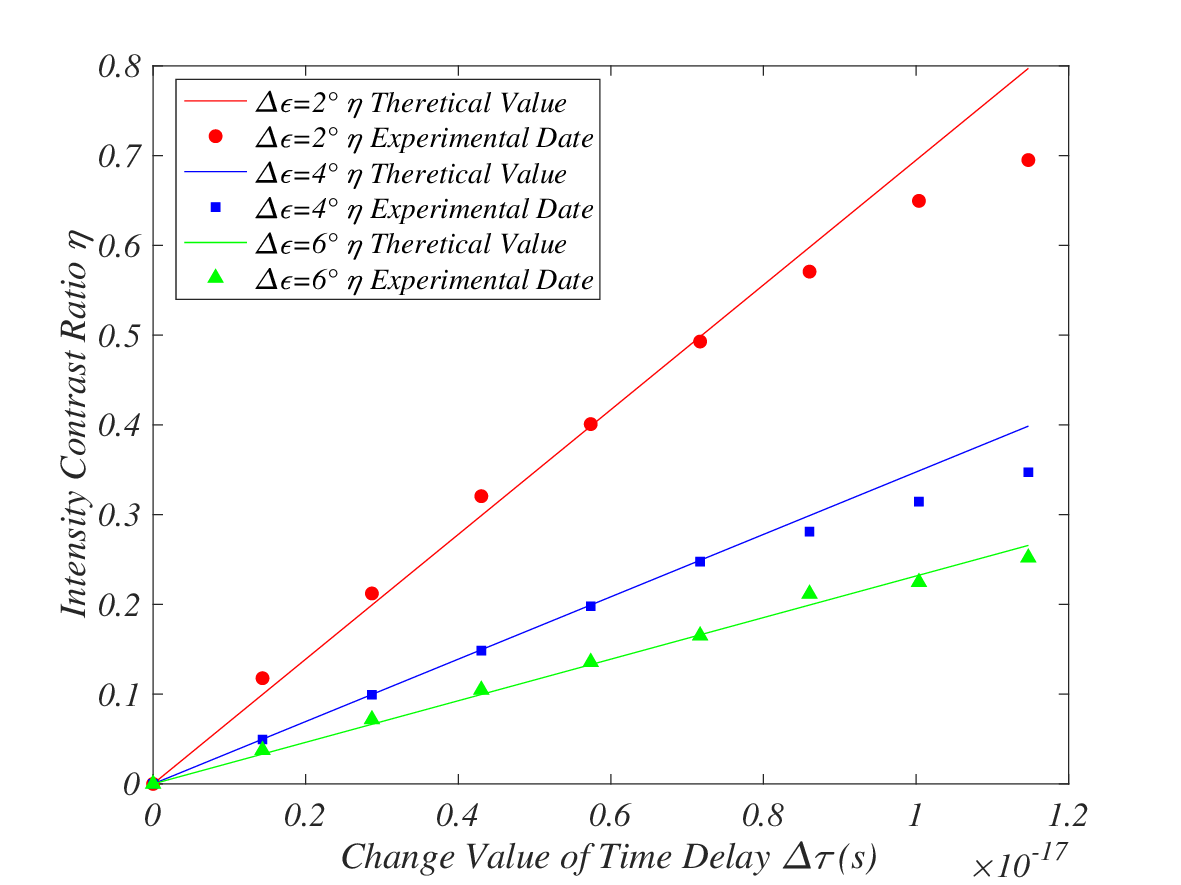}};
        \node at (0.4,5.5) {\textbf{(b)}};
    \end{tikzpicture}
    \caption{The distributions of QFI $Q_{WVA}$, intensity contrast ratio $\eta$, and shift of wavelength $\Delta\lambda$ under small and large time delay conditions. (a), the black solid line represents the intensity contrast $\eta$, the red dashed line represents the QFI distribution $Q_{WVA}$ after post-selection, and the blue dashed line represents the QFI $Q_{j}$ before post-selection. (b), the black solid line represents the wavelength shift $\Delta\lambda$, with the red and blue dashed lines matching those in (a). The time delay range between small and large delays differs by $10^{-12}$ orders of magnitude.}
    \label{fig:5}
\end{figure}

In the experiment, post-selection angles $\Delta \epsilon$ were set to $2^{\circ}$ (red), $4^{\circ}$ (blue), and $6^{\circ}$ (green), as shown in Fig. 5(a). The solid lines represent the initial optical power $I_{0}$ without disturbance, while the dashed lines show the post-selection optical power $I_{1}$ as a function of time delay variations $\Delta \tau$. It is evident that smaller post-selection angles $\Delta \epsilon$ lead to lower post-selection optical powers $I_{0}$ and $I_{1}$, consistent with theoretical predictions for post-selection probabilities.

Fig. 5(b) illustrates the response of the intensity contrast ratio $\eta$ to the time delay variations $\Delta \tau$ for different post-selection angles $\Delta \epsilon$. The solid lines represent theoretical simulations under conditions $\omega_{0} \Delta \tau <<\Delta \epsilon <<1$ , $\sigma^{2} \tau^{2} <<1$, showing a linear relationship $\eta \approx 2\omega_{0} \Delta \tau /\Delta \epsilon$, which aligns well with the experimental results. The intensity contrast $\eta$ is proportional to the sensing time delay variations $\Delta \tau$. Theoretically, smaller post-selection angles would result in a larger response of the intensity contrast ratio $\eta$ per unit sensing time delay variations $\Delta \tau$, leading to higher sensitivity for detecting $\Delta \tau$. In fact, for the monochromatic light source used in our system, very high measurement accuracy and sensitivity can be maintained over longer sensing fibers, provided that the post-selection angle is appropriately modulated.

\subsection{Water pressure sensing experiment}

We demonstrate a water pressure sensing experiment using the proposed WM scheme, with the setup shown in Fig. 6. 
\begin{figure}[t]
    \centering
\includegraphics[width=\linewidth,keepaspectratio]{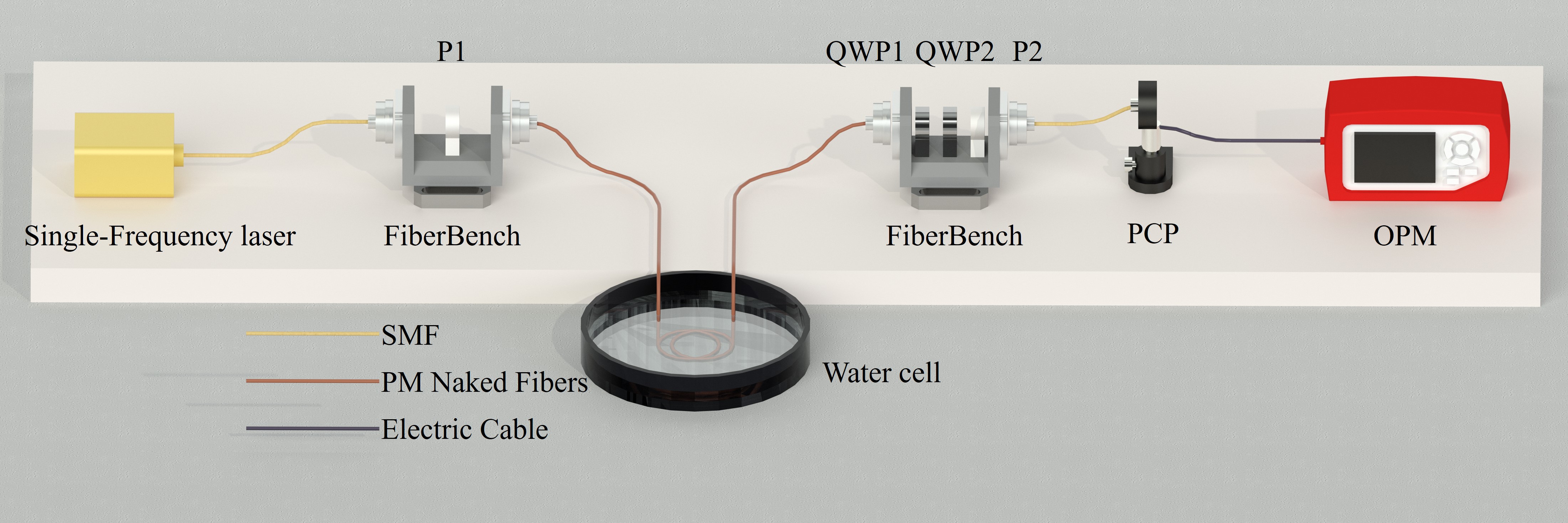}
    \caption{Water pressure sensing setup. The sensing optical fiber is placed in a water cell, where the weak interaction is induced by the water pressure applied around the fiber.}
    \label{fig:8}
\end{figure}

The light source, pre-selection, post-selection, and detection equipment are the same as in the principle verification experiment. The weak interaction occurs by partially fixing the sensing fiber at the bottom of a water cell, where it interacts with the water environment. Variations in the water volume apply pressure to the fiber, causing small disturbances that induce time delay variations. These variations propagate through the fiber and are detected, enabling water pressure sensing. The water pressure shows a linear relationship with the injected water volume,

\begin{equation}
\Delta P=\frac{\Delta V}{\pi R^{2}}.
\end{equation}

The radius $R=5cm$ of the bottom of the water cell, along with a water volume change of $\Delta V=20ml$, ensures a corresponding change in water pressure of $\Delta P=25.5Pa$. Under the influence of this water pressure, the PMF deforms, inducing a phase difference $\Delta \varphi$ between the two orthogonal polarization modes. The relationship between the water pressure and the phase difference is given by:

\begin{equation}
\Delta \varphi =\left\{ 2\frac{\beta \mu}{E} +\frac{\beta n^{2}}{2E} \left[ \left( 1-\mu \right) \left( p_{11}+p_{23} \right) -2\mu p_{12} \right] \right\} /d\cdot \Delta P,
\end{equation}
where $\beta =1.446\times 10^{7}/m$ represents the propagation constant, $E=6.4\times 10^{10}N/m$ denotes the Young's modulus, $\mu =0.2$ is the Poisson ratio, $n=1.454$ refers to the effective birefringence of the sensing optical fiber, and $p_{11}=0.121$ and $p_{12}=0.27$ are the Pockels coefficients of the sensing optical fiber, $d$ is the effective sensing fiber length subjected to water pressure.

In the experiment, we placed a $d =0.2m$ long sensing optical fiber at the bottom of the water cell. Each time, the pressure was increased by $25.5 Pa$, theoretically causing a phase shift of $\Delta \varphi =7.1\times 10^{-4}rad$, which corresponds to a time delay change of $\Delta \tau =5.83\times 10^{-19}s$. With the post-selection angle fixed at $5.5^{\circ}$, we monitored the continuous light intensity and intensity contrast ratio changes during water injection, as shown in Fig. 9. The graph of intensity contrast ratio versus injected water volume reveals a discrepancy between theoretical and experimental results, particularly after 50 ml of water is added, where the difference becomes more pronounced. However, the overall trend is consistent, and the sensing sensitivity matches the expected values from theoretical validation.

\begin{figure}[ht]
    \centering
    \begin{tikzpicture}
        \node[anchor=south west,inner sep=0] (image1) at (0,0) {\includegraphics[width=8cm,height=6cm]{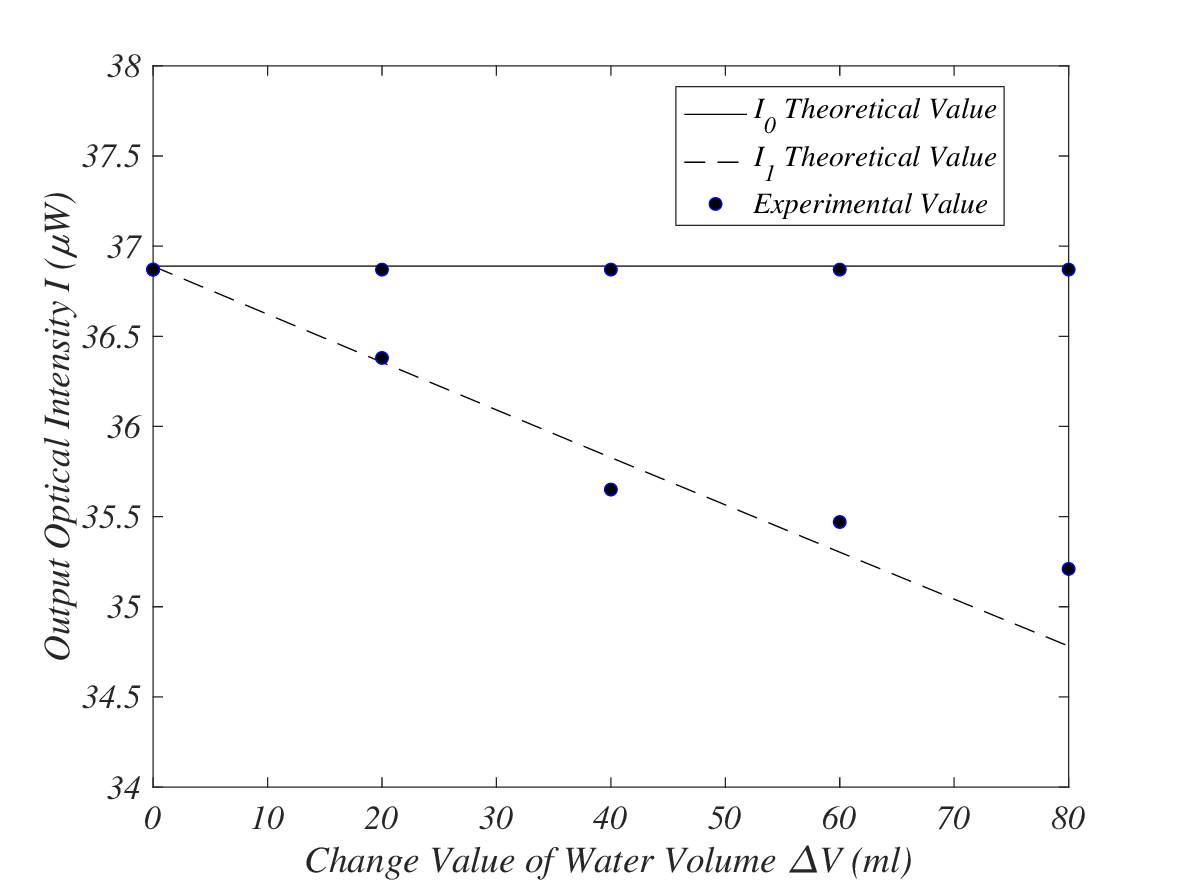}};
        \node at (0.3,5.5) {\textbf{(a)}};
    \end{tikzpicture}
    \hspace{0.5cm} 
    \begin{tikzpicture}
        \node[anchor=south west,inner sep=0] (image2) at (0,0) {\includegraphics[width=8cm,height=6cm]{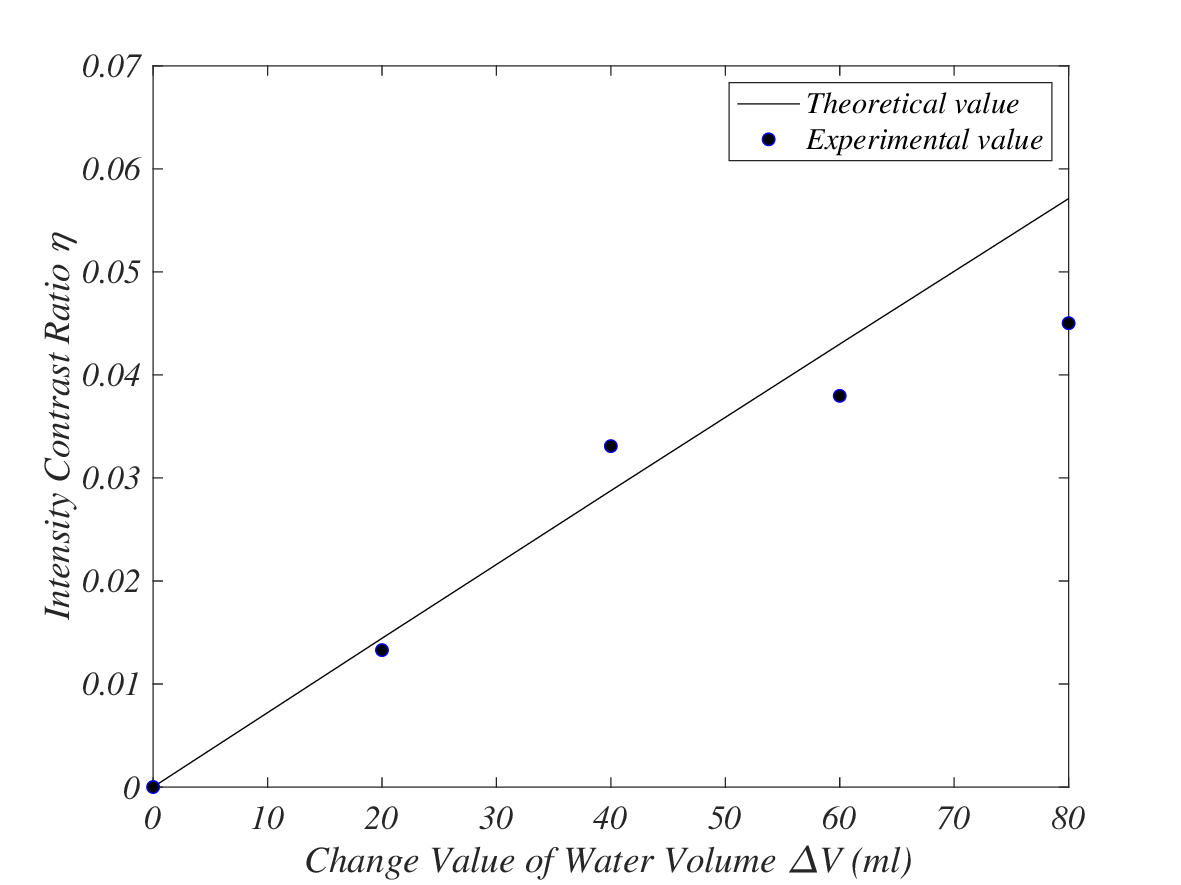}};
        \node at (0.3,5.5) {\textbf{(b)}};
    \end{tikzpicture}
    \caption{Water pressure sensing experimental results with a post-selection angle of 5.5°. For each 20 ml of water injected, the corresponding change in pressure is 25.5 Pa. Five consecutive injections were made. (a) shows the variation in post-selected optical powers $I_0$ and $I_1$ with the amount of water injected. (b) shows the change in intensity contrast with the amount of water injected, with the theoretical values indicated by the black curve.}
    \label{fig:10}
\end{figure}

The primary reason for this discrepancy is the presence of multiple environmental variables affecting the sensing fiber. First, the sensing fiber used is a silica-based panda-eye optical fiber, which is sensitive to temperature fluctuations \cite{Chen04072022}. Additionally, each time water is injected into the water cell, fluctuations in the water cause slight movements in the sensing fiber, leading to unwanted vibrations. These factors introduce noise during the optical power measurement process.

\subsection{Fiber alignment noise analysis}
In the experiment, the modular nature of the fiber optic system required connectors between components. Polarization-maintaining connectors are typically used for PMF devices to ensure proper alignment of the birefringent axes. However, due to manufacturing variations, the angle between the fast and slow axes and the ferrule is not fixed, leading to small alignment errors. These errors cause a rotation of the initial polarization state of the prepared light beam, which can impact the performance of the proposed scheme.

\begin{figure}[t]
    \centering
\includegraphics[width=\linewidth,keepaspectratio]{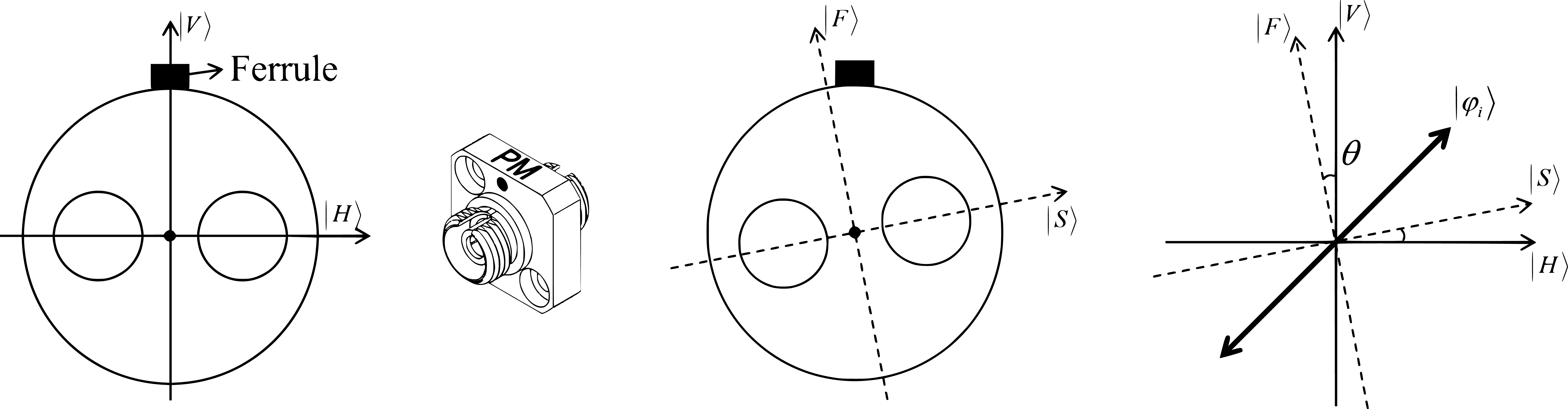}
    \caption{Principle of noise generation due to fiber alignment errors. At the fiber connector junction of the PMF, the ferrule is aligned along the direction perpendicular to the slow axis of the PMF, i.e., the direction labeled as $|V\rangle$ in the figure. In optical fiber systems, direct fiber connections are unavoidable, especially for PMF. However, due to manufacturing process variations, there is an alignment error angle $\theta$ between the ferrule and the slow axis of the PMF when using polarization-maintaining flanges for connection. This alignment error angle affects the polarization transmission, introducing alignment-related errors in the system.}
    \label{fig:4}
\end{figure}

In the fiber-based WM setup, perfect alignment between the polarization basis vectors ($|H\rangle$ and $|V\rangle$) of the incident light and the fast and slow axes of the sensing PMF cannot be ensured. Misalignment between these directions introduces noise, as shown in Fig. 8. While the pre-selected polarization states $H$ and $V$ can initially align with the PMF axes, fiber connectors inevitably introduce alignment errors. Specifically, the pre-selected polarization basis vectors misalign with the PMF axes by an angle $\theta$. A coordinate transformation is used to quantify this misalignment and express the polarization direction of the incident light.

\begin{equation}
|H\rangle =\sin \theta |S\rangle +\cos \theta |F\rangle,
\end{equation} 
\begin{equation}
|V\rangle =\cos \theta |S\rangle -\sin \theta |F\rangle,
\end{equation} 
where $|S\rangle$ and $|F\rangle$ denote the slow and fast axis directions of the PMF, respectively. When the incident linearly polarized light is oriented at a $45^{\circ}$ angle relative to the polarization directions of the fast and slow axes of the PMF, the polarization state of the incident light, denoted as $|\varphi_{i} \rangle =\frac{1}{\sqrt{2}} \left( |H\rangle +|V\rangle \right)$, enters the sensing PMF. The phase difference $2\omega \tau$ introduced between the fast and slow axes of the sensing fiber corresponds to a unitary evolution $\hat{U} =e^{-i\tau \omega \hat{A}}$ of the beam polarization state.

After undergoing evolution and post-selection through the WM process, the final post-selection success probability is:
\begin{equation}
\begin{aligned}
    P_{f}^{\prime} = \int \bigg\{ 
    &\Big[ \cos \theta \cdot \sin \theta \cdot \sin \left( \omega \tau - \varepsilon \right) \\
    &- \cos \theta \cdot \sin \theta \cdot \sin \left( -\omega \tau + \varepsilon \right) \Big]^{2} f^{2}(\omega) \\
    &+ \Big[ \cos^{2} \theta \cdot \sin \left( -\omega \tau + \varepsilon \right) \\
    &+ \sin^{2} \theta \cdot \sin \left( \omega \tau + \varepsilon \right) \Big] f^{2}(\omega) 
    \bigg\} d\omega.
\end{aligned}
\end{equation}

To simulate the post-selection success probability $P_{f}^{\prime}$ and intensity contrast ratio $\eta$ under different alignment error angles ($0.1 rad$, $0.3 rad$, $0.5 rad$) in the presence of large intrinsic delay, we use a monochromatic light source with a central wavelength of 1550$nm$ and a linewidth of 3.7$kHz$. The results are compared with the ideal case where no alignment error is present. The simulation results are shown in Fig. 9.

\begin{figure}[h]
    \centering
    \begin{tikzpicture}
        \node[anchor=south west,inner sep=0] (image1) at (0,0) {\includegraphics[width=8cm,height=6cm]{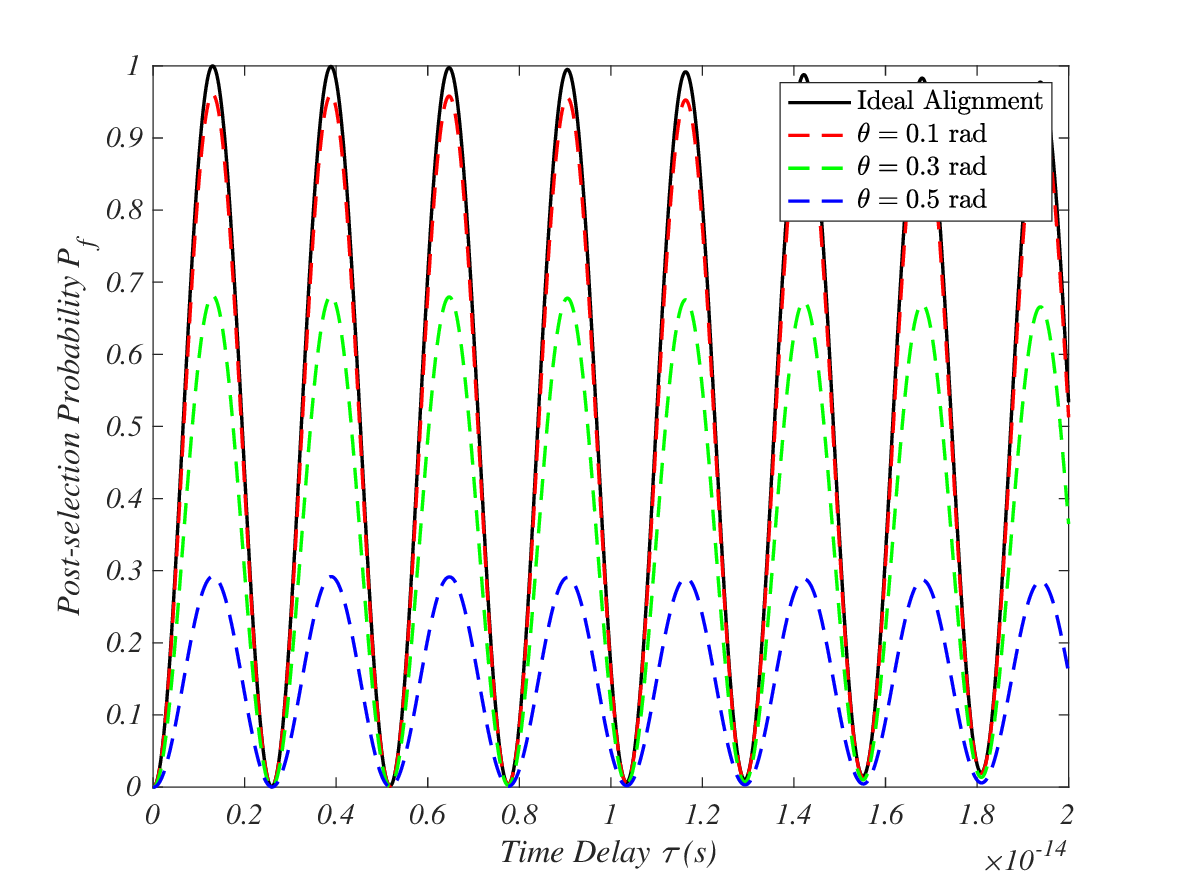}};
        \node at (0.4,5.5) {\textbf{(a)}};
    \end{tikzpicture}
    \begin{tikzpicture}
        \node[anchor=south west,inner sep=0] (image2) at (0,0) {\includegraphics[width=8cm,height=6cm]{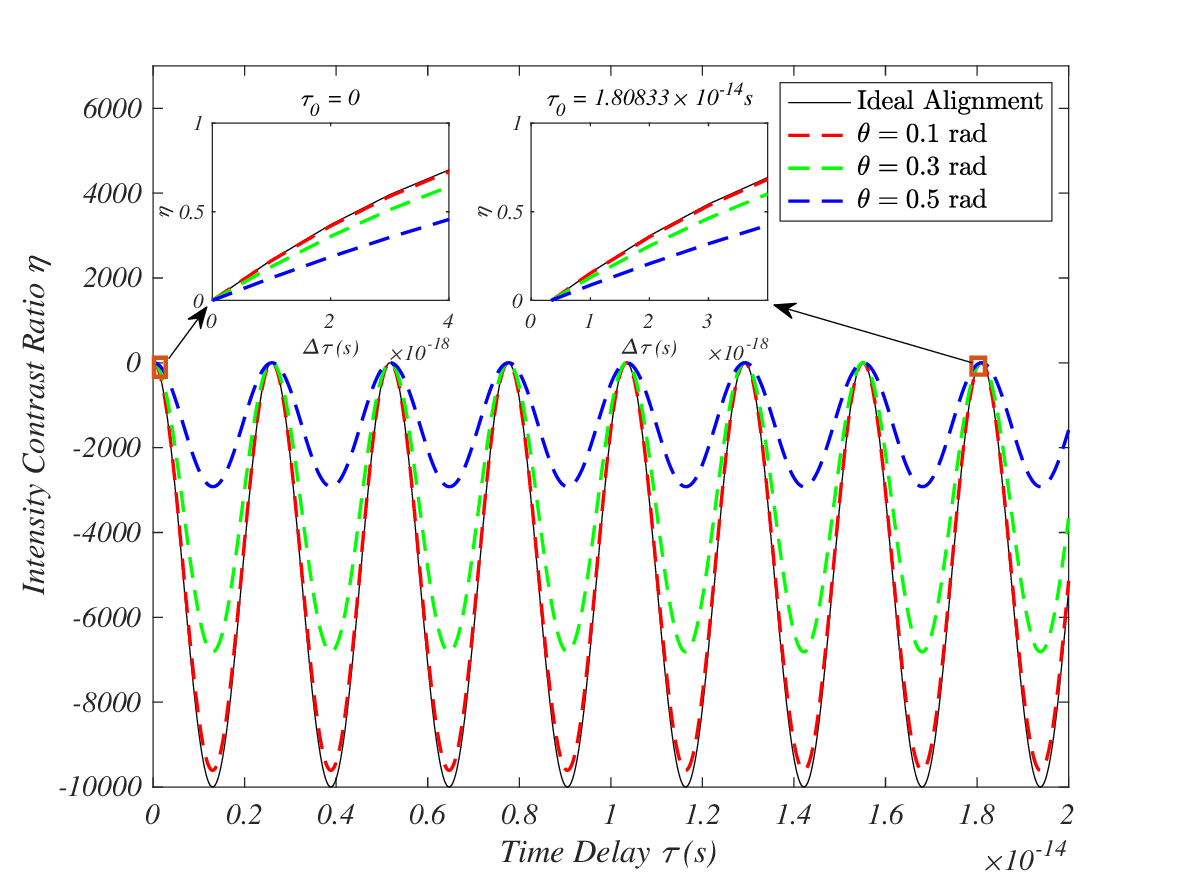}};
        \node at (0.4,5.5) {\textbf{(b)}};
    \end{tikzpicture}
    \caption{The distribution of post-selection success probability $P_{f}^{\prime}$ and intensity contrast ratio $\eta$ as functions of large delay time $\tau$ for different alignment error angles $\theta$. The black solid line represents the ideal case with no alignment error, while the red, green, and blue lines correspond to $\theta=0.1rad$, $\theta=0.3 rad$, and $\theta=0.5 rad$, respectively. (a) shows the variation of $ P_{f}^{\prime}$ for different error angles. (b) illustrates the variation of $\eta$ with $\tau$, with an expanded view of the linear regions ($\tau_{0} =0$ and $\tau_{0} =1.8033\times 10^{-14}s$) where the delay times significantly differ.}
    \label{fig:9}
\end{figure}

From Fig. 9(a) about the post-selection success probability distribution in the large inherent time delays range, the black solid line represents the ideal alignment scenario $\theta=0$, where the post-selection success probability $ P_{f}^{\prime}$ can reach 1. However, in practical experiments, due to the limited extinction ratio of various devices, $ P_{f}^{\prime}$ can only approach 1. The red dashed line corresponds to the alignment error angle $\theta=0.1rad$, approximately $5.73^{\circ}$, which nearly overlaps with the ideal alignment case, but the peak value $ P_{f}^{\prime}$ is reduced. The green and blue lines represent cases $\theta=0.3 rad$, and $\theta=0.5 rad$, respectively. As the alignment error angle increases, the peak value $ P_{f}^{\prime}$ gradually decreases, which will directly affect the distribution of intensity contrast ratio $\eta$ in the large inherent time delays range.

Fig. 9(b) shows the simulation results of the intensity contrast ratio $\eta$ distribution in the large inherent time delays range. The curves in different colors have the same meaning as in FIG. 5(a). The two subplots represent the situations for small time delay $\tau_{0}=0$ and large inherent time delays $\tau_{0} =1.8\times 10^{-14}s$. It is clear that the intensity contrast ratio $\eta$ does not change significantly with time delay variations $\Delta \tau$. However, as the misalignment error angle $\theta$ increases, particularly when it reaches 0.3$rad$ or even 0.5$rad$, the intensity contrast ratio $\eta$ shows a noticeable decrease in responsiveness, becoming more gradual both for small and large inherent time delays scenarios. This leads to a significant reduction in sensitivity,  and the results remain consistent for small and large inherent time delays. In the current fiber optic device manufacturing industry, the alignment error required by international standards is much smaller than the misalignment angles we have analyzed \cite{IEC2020}. Therefore, within a reasonable range of alignment errors, this scheme demonstrates a strong resistance to noise caused by fiber misalignment.

\section{conclusion}
\label{sec:IV}
In this work, we propose a WM-based scheme for high-precision fiber-optic delay estimation under large inherent time delays, addressing the decoherence problem caused by significant inherent time delays between two orthogonal polarization modes in long-distance fiber transmission. By utilizing intensity contrast signals, we define a signal that is adjusted based on the post-selection angle, which is controlled by adjusting the optical power at the fiber output. We demonstrate that narrower linewidths of the light source result in a longer effective transmission range for high-sensitivity signals, even in the presence of large inherent time delays caused by polarization mode dispersion. This ensures measurement accuracy and sensitivity comparable to BWM. 

Experimental results confirm the feasibility of the proposed scheme, showing that after post-selection through weak measurement, the output optical intensity is highly sensitive to small time delay variations between the two orthogonal polarization modes propagating through the fiber. This allows for the detection of time delay changes with a precision exceeding the theoretical limits of the scheme, as evidenced by a detection sensitivity of 25.5 Pa in water pressure changes. Additionally, we theoretically correct for alignment error noise in the fiber-optic system, showing that our scheme exhibits robustness against small alignment errors, thus maintaining high precision even under such conditions.

Given its long-range capabilities, structural simplicity, and robustness to device imperfections \cite{PhysRevLett.107.113603}, our scheme holds significant potential for practical applications in high-precision fiber-optic sensing and offers a new approach to long-distance distributed fiber-optic sensing.

%\appendices


\section*{Acknowledgment}
This work is supported by National Natural Science Foundation of China ( 62371199, 62071186 ); Natural Science Foundation of Guangdong Province (2024A1515012427), Guangdong Provincial Quantum Science Strategic Initiative (GDZX2305001), Guangdong Provincial Key Laboratory (2020B1212060066). 
\bibliographystyle{unsrt}
\bibliography{Refference}

\end{document}